# Unique Scales Preserve Self-Similar Integrate-and-Fire Functionality of Neuronal Clusters

Anar Amgalan[1†], Patrick Taylor[2†], Lilianne R. Mujica-Parodi[1,3], and Hava T. Siegelmann[2,4]

[†]Authors contributed equally to the submitted work.

1. Laufer Center for Physical and Quantitative Biology,
   Physics and Astronomy Department and
   Department of Biomedical Engineering
   Stony Brook University, Stony Brook NY—USA

2. College of Information and Computer Sciences
   University of Massachusetts, Amherst MA—USA

3. Athinoula A. Martinos Center for Biomedical Imaging
   Department of Radiology
   Massachusetts General Hospital/Harvard Medical School, Charlestown MA—USA

4. Neuroscience and Behavior Program
   University of Massachusetts, Amherst MA—USA

Please address all correspondence to:
Lilianne R. Mujica-Parodi, Ph.D.
Professor and Director, Laboratory for Computational Neurodiagnostics
Department of Biomedical Engineering
Stony Brook University
Phone: 631-428-8461
Email: lilianne.strey@stonybrook.edu
LAB WEBSITE: www.lcneuro.org

**Identifying the brain's neuronal cluster size to be presented as nodes in a network computation is critical to both neuroscience and artificial intelligence, as these define the cognitive blocks required for building intelligent computation[1]. Experiments support many forms and sizes of neural clustering[2-14], while neural mass models (NMM) assume scale-invariant functionality[15]. Here, we use computational simulations with brain-derived fMRI network[16,17] to show that not only brain network stays structurally self-similar continuously across scales, but also neuron-like signal integration functionality is preserved at particular scales. As such, we propose a coarse-graining of network of neurons to ensemble-nodes, with multiple spikes making up its ensemble-spike, and time re-scaling factor defining its ensemble-time step. The fractal-like spatiotemporal structure and function that emerge permit strategic choice in bridging across experimental scales for computational modeling, while also suggesting regulatory constraints on developmental and/or evolutionary "growth spurts" in brain size[18-20], as per punctuated equilibrium theories in evolutionary biology[21,22].**

Introduction

Neuronal clustering is observed far more frequently than by chance [2]. The most common types of neuronal clustering believed to play a role in neuronal activity, development, and modularity [3,4] are *dendritic bundles* (neurons clustered together at their apical ends, with axons terminating at the same target [5]), *minicolumns* (radially clustered cell bodies, often around 80-100 neurons), *columns* (radially grouped minicolumns, often around 60-80 minicolumns [6,7]), *cluster-columns* (single minicolumn surrounded by one circular layer of columns, often seven or eight columns total [8], also defined by their functional role: e.g. color processing), as well as small *lego blocks* (of one order of magnitude only) [2,9-11]. Similarly, temporal clustering of spikes is observed more frequently than predicted by spiking rates alone [12,13 14,]. Yet the brain's "least functional unit" above the neuron, or *monad* is still undefined. The potential utility of defining a monad is that the "least functional unit" may actually be a "sufficient functional unit." In practical terms, identifying neural monads tells us which modules are cognitive or constitute computational functions worth modeling. This is of direct relevance for choosing the most strategic scales to measure and model in computational neuroscience, as well as for building intelligent computation [1].

The diverse experimental modalities utilized in brain sciences probe scales that can overlap little and obtain information through different physiological aspects of brain. Microscopic techniques resolve individual neurons, but lack coverage; fMRI captures an image of the entire cortex, but at a resolution of million or so neurons per voxel; EEG and NIRS are limited by depth and coarse spatial resolution, electrode readings can reach spatial extent of >1mm of tissue, but miss on the biophysical details of individual neurons. Here, through our computational coarse-graining of spiking network activity data, we provide justification for a procedure of joining distinct scales, and therefore information from some of the distinct imaging modalities available for a fuller picture of brain's workings. We establish an instance of scale-invariance of spiking network's functionality at the transition from a single neuron to an ensemble of handful of neurons and suggest that a similar procedure can detect a pair of experimental scales and techniques (e.g. calcium imaging combined with multi-electrode array recording) that are amenable to simultaneous modeling by virtue of sharing functional organizing principle. Such an estimate of paired similarly-behaving scaled would guide the choice of imaging modalities to inform multi-scale modeling efforts aimed at maximal coverage of brain's dynamical repertoire.

Some understanding of the dynamic repertoire of brain's larger than neuron structures, and particularly their repetitive and self-similar nature has been obtained: synfire chain [12,13] describes the sequentially wired sub-assemblies propagating synchronous activities; avalanche models describe the power-law statistics of sizes of events in multielectrode recordings [23]; power-law noise statistics in brain signals have been robustly characterized [24-26]. Also reliably documented is the scale-free structural properties of brain network [27]. Yet the rules unifying structures and processes in brain at distinct scales aren't fully described.

The most basic functionality of brain is that of spiking neurons. Neurons, however variable, act as non-linear summing integrators, with a decay in potential and a threshold for firing; they spike when excited frequently and strongly enough. Here, we propose a scale-invariant dynamic, where a pair of similar views of the same object are separated by a specific re-scaling factor that transforms one into another. A simple analogy is provided by Sierpinski triangle [28], where a finer stage is obtained from the previous coarser stage by halving the spatial dimension of each triangle and tripling the number of triangles, thereby characterizing each stage of re-scaling of the fractal object by two integers: (2, 3). Here, we propose a functional scale-invariance or "fractality," such that a redefinition of the spiking network's basic primitives: node, edge, and spike as a particular multiple of themselves allows one to observe the neuron-like signal-integration functionality re-emerge in the coarser structure, albeit only at unique re-scaling. The consequence of this proposition is that multi-scale models can coarse-grain for computational expediency while strategically choosing a unique set of scales specifically determined to permit translation. Thus, we expand on previous studies, which have demonstrated the feasibility of hierarchical clustering [29] and networks' scale-free features [30-32] with regard to brain *structure* to introduce self-similarity of *dynamics* of clusters of neurons, replicating integrate-and-fire-like functionality across scales.

To computationally test whether dynamics of signal transmission are preserved in a manner akin to "functional harmonics" of the original scale, we construct a spiking network that follows actual brain connectivity. The starting point of our study is an all-to-all resting-state human fMRI-derived functional connectivity matrix, extracted from 91,282 voxels, and providing full coverage of the human cortex [16,17]. The fMRI data describe the temporal correlation of each of these voxels with every other voxel, allowing an inference about the degree to which they are functionally connected. Such inferences have been externally validated by comparison to structural and DTI data [33]. Each voxel of this matrix is $(2mm)^3$, and thus measures a compensatory hemodynamic (blood oxygen level dependent, or BOLD) response across ~1,000 cortical minicolumns, or about ~1M neurons. Each voxel is represented as a node in our initial network, and in our computational experiment it follows the functionality of a leaky integrate-and-fire neuron. We start our computational experiment from an already non-neuronal scale of data acquisition, as we suggest that the observations made on fMRI scale may be close to one of the stages of the potential network re-scalings, along which the signal integration property is preserved.

The initial fine-grain network is reduced into clusters using a streaming hierarchical clustering algorithm [34]. We chose a streaming clustering algorithm for its computational expediency and reduced memory requirement. In deciding which nodes combine to form a cluster, we chose full-linkage clustering, requiring that only sets of nodes that have all-to-all connection weights higher than the threshold *c* become a cluster of nodes (*an ensemble-node*).

All pairs of neurons $i, j$ belonging to the same cluster satisfy: $w_{ij} \geq c$, where $w_{ij}$ is the connection weight between nodes $i$ and $j$. By adjusting threshold $c$ one can flexibly control the average number of neurons in an ensemble-node. This clustering procedure allows re-scaling to a continuous spectrum of size of networks (**Fig. 1A-C**).

The set of edges running between two neighboring ensemble-nodes aggregate to give the *ensemble-edge* connecting these ensemble-nodes. The weight of the ensemble-edge is given by the mean weight of the edges defining it: $w_{kl}^{cg} = mean(w_{ij})$ where $i$ runs over all neurons in cluster $k$ and $j$ runs over all neurons in cluster $l$. For each ensemble-node we now define the *node strength* as the sum of weights of edges connecting to it: $s_k^{cg} = \sum_l w_{kl}^{cg}$, where superscript $cg$ indicates that quantity pertains to the coarse-grained network.

We next redefine the activity of the coarse-grained network by temporally clustering firings of an ensemble-node into an *ensemble-spike*. An ensemble-node consisting of $N_N$ nodes produces an ensemble-spike if its nodes produce a combined burst of above-threshold ($N_S$) number of spikes in the time-step, and no ensemble-spike occurs otherwise. This characterizes the number of ensemble-spikes required to arrive in rapid succession into an ensemble-node prior to activation of its own ensemble-spike. Intuitively, only areas dense enough in spikes in fine level spike raster (**Fig. 1D**) become ensemble-spikes in the coarse network's ensemble-spike raster (**Fig. 1E**) in the corresponding ensemble-node and time step. A spiking neuron's refractory period also re-emerges in the coarse-grain network, and concomitantly guides our choice of *ensemble-step*: the time step re-scaling parameter. Further details in the **Methods Section**.

Results

***Coarse-grained structure is self-similar.*** Coarse-graining at various cutoffs produces networks with a range of ensemble-node sizes and number of ensemble-nodes **(Fig. 2A)**. Importantly, cutoff has little impact on the tail of ensemble-nodes' size distribution, which is invariant with respect to mean count and power-law behavior, suggesting self-similarity in agreement with a body of previous work [27,35,36] **(Fig. 2B)**. The original functional brain network exhibits a self-similar property when it comes to its new coarsened structure: the connectivity network preserves the geometry of its node strength distribution after the coarse-graining procedure **(Fig. 2C)**.

***Coarse-grained dynamics are self-similar.*** Re-scaling our network to a spectrum of average ensemble-node sizes ($N_N$) reveals signature of an integrate-and-fire property to a varying degree depending on the re-scaling factors ($N_N$ and $N_S$) **(Fig. 3A, *solid lines*)**, but with three key features. Firstly, leading up to an outgoing ensemble-spike, the cross-correlogram of ensemble-spiking activity weighted by the ensemble-edge weights (see Methods for full description) displays integration of a faster-than-exponentially rising amount of inputs from neighbors. Secondly, the incoming inputs peak at the time-step immediately preceding ensemble-spike output. Finally, incoming activity from a given ensemble-node's neighbors abruptly drops at the time of an outgoing ensemble-spike.

***Self-similar dynamics are preserved only at specific scaling factors displaying "functional harmonics" of the original scale.*** This scale invariant dynamics results from the brain's organizational structure. Comparing the sharp increase in signal integration preceding the ensemble-spike observed in the clustered network to that of a control network that was coarse grained to an identical average size of ensemble-nodes but through random clustering (ignoring

the edge weights), we see a control behavior that is only exponential **(Fig. 3A, dashed lines)**, indicating its weak input integration. Shaded areas in panels of **Fig. 3A** indicate the extent of the coarse network's input integration beyond that of an equivalently sized randomly clustered network. The top panel (ensemble-spike requiring >=2 spikes) shows little difference from the random case, while the bottom panel (ensemble-spike requiring >=4 spikes) shows that inputs of >=4 spikes are integrated much more rapidly leading up to an ensemble-spike (solid line) than in a randomly clustered network (dashed line). From the starting network of 7,784 nodes, by searching the space of re-scaling factors and maximizing the *integration coefficient* (see **Methods section** for details), we find ensemble-node size of ~10 nodes and ensemble-spike size of ~5 spikes to maximally recover integration of inputs leading up to ensemble-spikes **(Fig. 3B left)**. We also establish the time-axis re-scaling factor, or ensemble-step, of ~4 by maximizing the refractoriness present in auto-correlogram **(Fig. 3C)**. Thus, the network coarse-graining procedure returns a unique triplet consisting of ensemble-node=10 nodes, ensemble-spike=5 spikes, and ensemble-step=4 that recovers properties of spiking nodes in a network of ensemble-nodes. We also observe that the nodes' local clustering positively correlates with the integration coefficient. Thus, ensemble-nodes with greater local clustering coefficient display greater signal-integration behavior, and the dependence does not deviate from a linear trend **(Fig. 3D)**.

## Discussion

Our fMRI-driven interrogation of simulated spiking network data suggests that brain organization has a dense multi-scale property in structure and functionality. Neurons naturally cluster across many scales, and such clusters create new networks with fewer functional units, but similar algorithmic features. Our coarse-graining procedure attempts to capture only one such transition between a pair of scales by detecting similarity of dynamics, but suggests that the brain potentially preserves its dynamics across scales at many distinct levels, recycling already existing functional building blocks. It is therefore feasible that brain's least functional unit larger than a single neuron might be a cluster of neurons (an "ensemble-neuron" in our terminology) displaying a neuron-like integrate-and-fire behavior. Whether neuronal clusters act as integrating neurons and if so, at what re-scaling factor, can be established by a targeted experiment of large-scale calcium imaging and other optical recording modalities. Such an experiment would, having simultaneously recorded from large number of neurons, apply the machinery we describe, identifying a pair of coarse and fine scales that display the most self-similar behavior. Such a constraint on the similarly behaving pair of scales would drastically narrow the space of multi-scale models of brain dynamics.

The implications of our work are particularly important for neural mass models (NMMs). We add to the existing experimental and analytical motivations for NMMs [37-42] and propose a mechanism for constraining the scales to be modeled with NMMs. NMMs provide an attractive procedure for reducing brain dynamics to just a few quantities at population level. Further supporting the treatment of neuronal systems as NMMs, we are providing computational justification for approximating the fine-scale neural dynamics (namely integrate-and-fire rule) by a similar dynamical rule, but at a particular coarser scale out of all scales empirically available for imaging. One implication of such approximation is that brain dynamics may be modeled using experimental data from several non-overlapping scales, as long as those scales are chosen strategically to retain self-similar dynamics. In addition, our analysis coarse-grains a network of neurons while preserving integrate-and-fire dynamics at network sizes exceeding 1000 neurons and ensemble-neurons, highlighting potential of NMM-like models to describe interactions of large number of neuronal populations. From a biological perspective, the periodicity of scale invariance may offer clues into developmental and evolutionary processes. The dynamic structure of the action potential (gating, excitation, inhibition, return to baseline) reflects, at its very basis, a mechanistic solution to the biologically ubiquitous problem of allostatic regulation in response to noisy inputs. In growing the brain from unicellular to multicellular to agglomerate structure—while continuing to maintain allostatic regulation—it therefore makes sense that dynamic signatures associated with such regulation would also be preserved. As such, the discontinuity of scales at which allostatic regulation continues to hold may impose key functional constraints on "growth spurts" in brain development[18-20], as per punctuated equilibrium theories in evolutionary biology[21,22].

## Methods

*Network structure.* The starting point of our study is an all-to-all resting-state fMRI-derived (N=68, repeated 4 times) functional connectivity matrix, extracted from 91,282 voxels, and providing full coverage of the human cortex [16,17]. Each voxel is 2mm$^3$, and thus measures a compensatory hemodynamic (*blood oxygen level dependent*) response across ~1,000 cortical minicolumns, or about ~1M neurons (in primates and most mammals, the cortex contains a 20/80 ratio of neurons to glia [45]). Each voxel is represented as a node in a graph.

*Nodes.* The nodes follow a leaky integrate-and-fire neuronal model, updated via change in membrane potential: $dV_k/dt = -\tau_v + \sum_j w_{jk} \times I_j + \sigma_k$ with spike conditions: if $V_k(t) > V_\theta$, then the node initiates a spike by setting: $V_k(t) = V_r$. Here:

$V_k(t)$ is the time-dependent membrane potential (voltage) of a neuron, with $k$ indexing the neuron integrating signal from its neighbors indexed by $j$.

| | |
|---|---|
| $I_j$ | indicates the input signal from pre-synaptic neurons, indexed by $j$. $I_j = 1$ during the time-step neuron $j$ fires and 0 otherwise. |
| $w_{jk}$ | is weight connecting the spiking neuron $j$ to the signal-integrating neuron $k$. |
| $V_r$ | is resting membrane potential, to which the neuron returns after spiking. |
| $V_\theta$ | is firing threshold for an action potential, upon reaching which the neuron initiates a spike. |
| $\tau_v$ | is rate of leakage in membrane potential from $V_k(t)$ towards $V_r$. |
| $\sigma_k$ | is the normally distributed noise input to neuron $k$. |

***Integration Coefficient and Normalization by Random Controls.*** We compare the well-clustered coarse network to a randomly clustered control to reveal effects of strong intra-cluster connectedness on coarse network's input integration. In order to quantify the effects of integration of incoming ensemble-spikes from neighbors in a coarse-grained network, we obtain the cross-correlogram of ensemble-spikes:

- For a particular ensemble-node, for each observed ensemble-spike, we collect the most recent ensemble-spike of each of its neighboring ensemble-nodes according to their time lag $\tau$, by which they preceded the ensemble-spike. For each time lag $\tau$, we then sum ensemble-edge weights for all ensemble-spikes that preceded our ensemble-spike of interest by $\tau$ time steps. Any neighboring ensemble-node contributes only one ensemble-spike to this list of sums – its most recent ensemble-spike preceding the ensemble-spike of interest, and to the sum of weights at observed time-lag $\tau$. This tells us how much information was integrated by the ensemble-node and from how many time steps prior to producing the ensemble-spike.

- For each time lag $\tau$, we then sum the ensemble-edge weights collected over all ensemble-spikes of all ensemble-nodes, to obtain the (ensemble-edge-weighted) cross-correlogram $P(\tau)$ dependent only on time-delay $\tau$. $P(\tau)$ tells us how much input (ensemble-spikes weighted by ensemble-edge weights) had to be received from neighboring ensemble-nodes $\tau$ time steps prior to a typical ensemble-spike in the coarse network.

- The cross-correlogram $P(\tau)$ is then normalized (divided) by the cross-correlogram $P_{random}(\tau)$ for the case if the original network was clustered randomly, and we integrate the ratio over interval of $\tau$, on which the ratio exceeds 1, weighing by the inverse of the time lag $\tau$ to discount inputs received early, at large $\tau$. We term this quantity *integration coefficient* ($IC$):

$$IC = \int_{P/P_{random}>1} \frac{1}{|\tau|} \frac{P(\tau)}{P_{random}(\tau)} d\tau$$

The integration coefficient then acts as a metric of the coarse network's similarity to a network of spiking nodes.

***Re-wired Controls***. As *node degree sequence* frequently drives emergent properties of a complex network [46], we next examine the role of isolating the fine-grain network's *node strength sequence* in determining whether integrate-and-fire behavior emerges at a coarser scale. *Node strength sequence* is the equivalent of degree sequence for a *weighted network*, and is given by the sum of weights of all edges of a given node. To quantify the contribution of node strength sequence to our observations, we provide a comparison with an original network shuffled while preserving each node's strength. The shuffling follows a local rewiring rule: given 4 nodes (a, b, c and d), edges ab, cd, ac, and bd are shuffled from $ab = cd = w_1$ and $ac = bd = w_2$ to:

ab = cd = $w_2$ and ac = bd = $w_1$. This preserves the local strength of node a as: ac + ab = $w_1 + w_2$ before and after the rewiring, while dissolving the network's cluster-forming topology. The value of integration coefficient is lower for a shuffled network than for original brain network **(Fig. 3B, right panel).** It is also notable that the shuffled network, when re-scaled, is unable to produce the same average cluster size at the same connection threshold as original brain network, nor to maintain ensemble-spikes consisting of large numbers of spikes **(Fig. 3B, right panel**, bottom right corner where data is absent – area indicated in crimson).

*Do clusters show neuron-like refractory periods?* By means of an auto-correlogram $P_{auto}(\tau)$ (instead of binning ensemble-spike inputs from neighboring ensemble-nodes, we bin the ensemble-node's own most recent ensemble-spike preceding the ensemble-spike of interest), we search for the time-axis re-scaling parameters that result in the strongest degree of refractory behavior, which we measure as the ratio of ensemble-spike auto-correlogram at $\tau = 2$ to that at $\tau = 1$: $P_{auto}(2)/P_{auto}(1)$. This helps us resolve whether a pair of consecutive ensemble-spikes of an ensemble-node had any systematic and noticeable time-gap in between, when ensemble-spiking was suppressed **(Fig. 3C)**.

*Further Variations.* A particular coarse-grain version of a spiking network is realized by choosing multiple parameters: (1) c (edge-weight cutoff for full-linkage clustering), which directly influences $N_N$ – average number of nodes in an ensemble-node, (2) $N_S$ – number of spikes in an ensemble-spike and (3) $N_T$ time-axis re-scaling factor. The latter is formally an integer factor, by which the time bin used for the coarse level statistics is smaller than the mean inter-spike interval of a single node of the fine-grain network. Re-scaling the time-axis so that the number of spikes in an ensemble-node expected from the neuronal spiking frequency alone is ~1 (i.e. setting $N_T=N_N$) permits singling out time bins with spiking activity elevated by a factor of $N_S$ as compared to the level expected from mean spiking frequency. These time-bins in the coarse ensemble-spike raster become the basis of the cross-correlogram and calculation of integration coefficient. Maximizing the integration coefficient with respect to $N_T$ after the combination of spatial ($N_N$) and activity ($N_S$) re-scaling parameters have been estimated, allows us to establish the full set of three parameters of re-scaling. This completes the analogy between spiking neurons and ensemble-spiking clusters of neurons. In search of the integrate-and-fire behavior for coarse network, we cover the space of spatial and activity re-scaling factors as follows: $N_N$ from minimum of ~3 neuron clusters up to ~12 in increments of 1; $N_S$ from minimal burst of two spikes to the maximum that can be sustained by the fine spiking network. The coarse-graining procedures described can also be iterated more than once, on already clustered network, by treating it as the fine-grain network and operating with the same machinery enabling a principled dynamics-preserving hierarchical reduction of functional models.

Figures and Captions

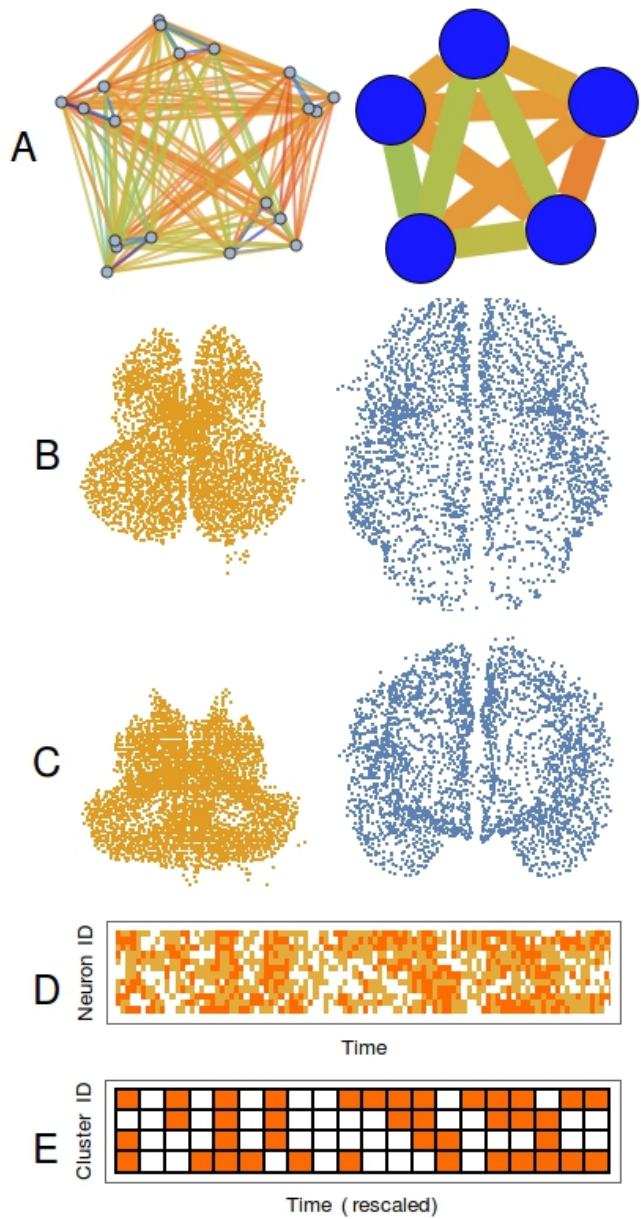

**Figure 1. Re-scaling procedure for neuronal clusters and spike clusters demonstrates how proximate regions cluster together, reproducing organ-scale brain morphology**. **(A)** We depict a fine-grain network of neurons *(left)* and its coarser, re-scaled version *(right)*. Each ensemble-node, at a coarser level, is a collection of nodes from finer resolution level, clustered based on their connection weights *(thin lines, left panel)* with all edges between a pair of clusters averaged *(thick lines, right panel)*. **(B)** Axial view of coarse-grained brain, after the procedure, at a cutoff of 0.98, re-scaling the original network of 91282 nodes down to 7,784 clusters. Each ensemble-node is located at the mean of coordinates of nodes comprising it. **(C)** Similarly, a coronal view. **(D)** The spike-train of fine-grain network is re-scaled into **(E)** ensemble-spike-train of coarser network.

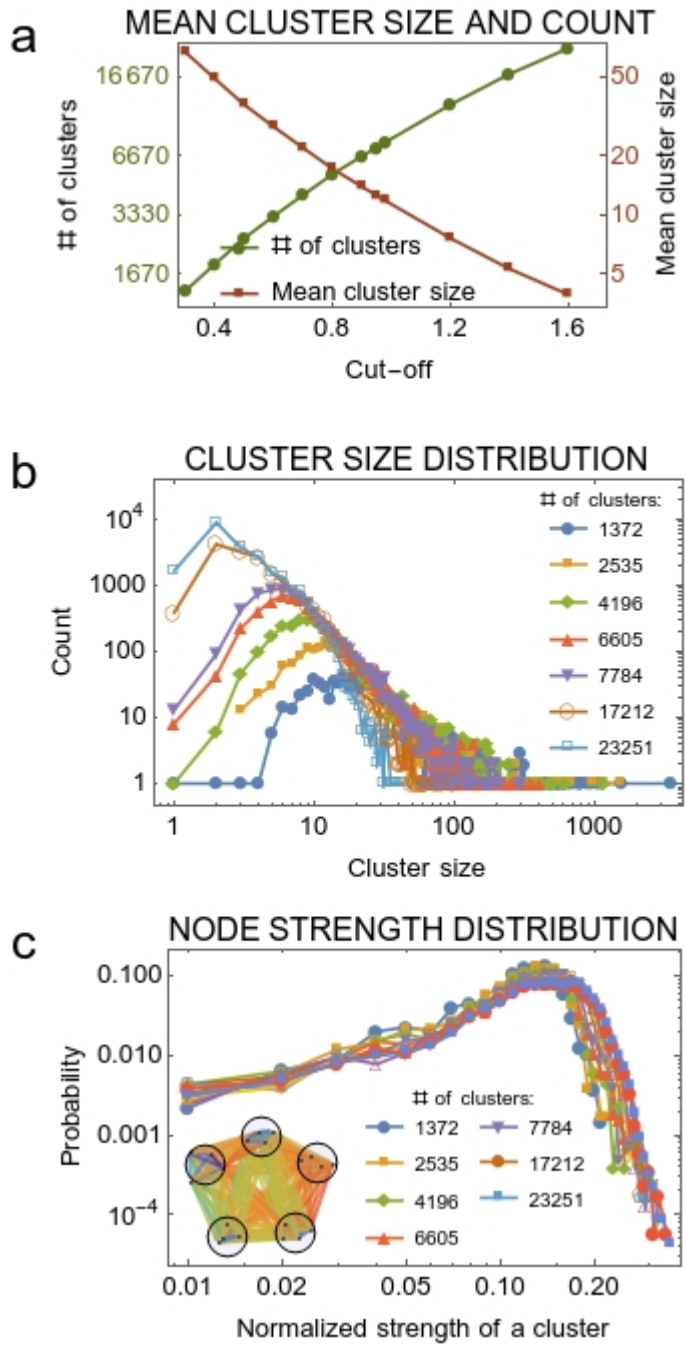

**Figure 2. The brain network naturally clusters into a spectrum of system sizes while preserving its degree distribution shape.** **(A)** Cutoffs are imposed on intra-cluster connection weights. **(B)** The network of 91,282 nodes is coarse-grained to obtain number of clusters ranging from ~1,300 to ~23,000, yet it retains a non-trivial most frequent cluster size and a power-law drop off in distribution of cluster sizes. **(C)** Node strength distribution of the network stays largely invariant when it is clustered to well-connected clusters (inset) and inter-cluster edge weights are averaged. We define node strength as sum of all edges of a node and treat as weighted network's version of "degree" in unweighted network.

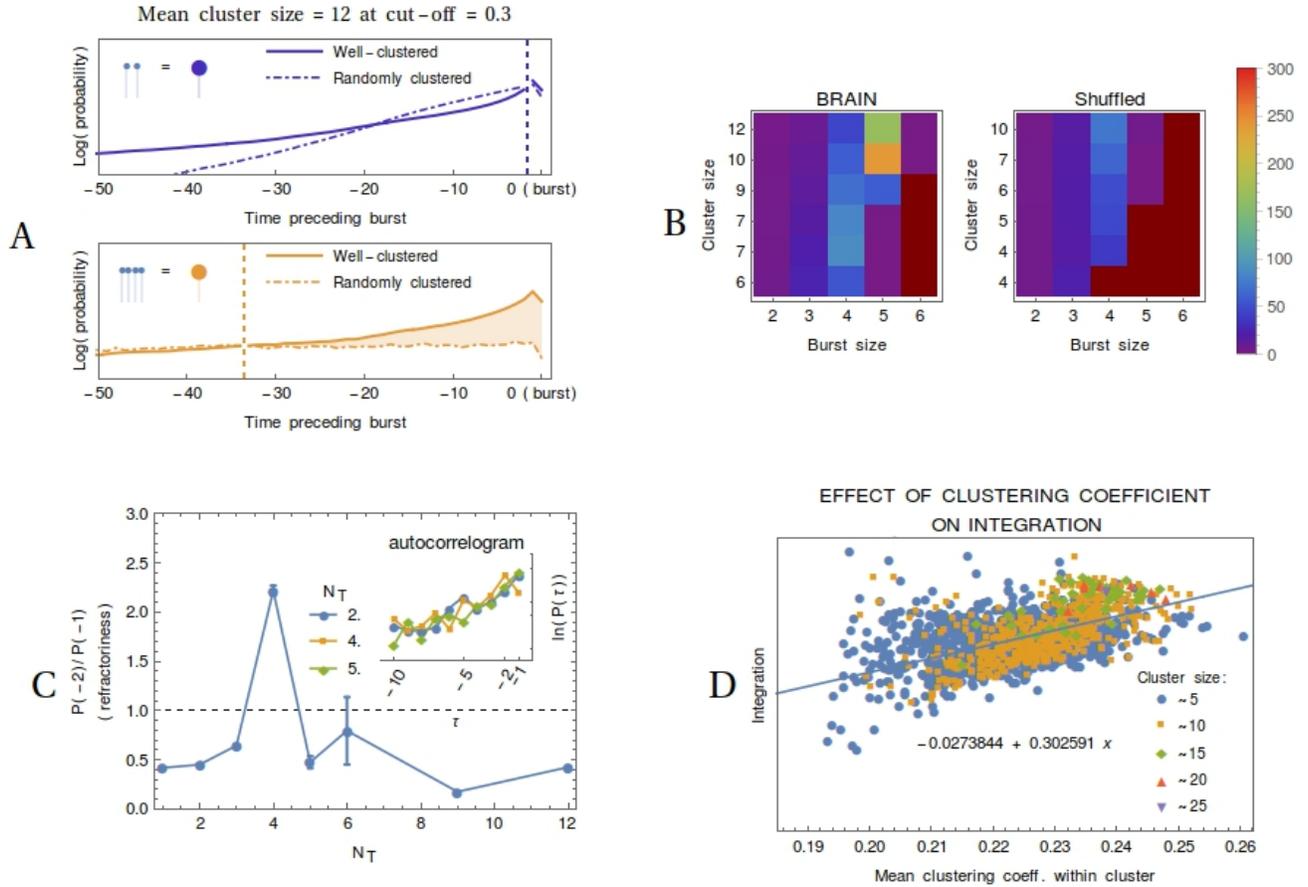

**Figure 3. Ensemble-edge-weighted cross-correlogram $P(\tau)$ of cluster-spiking behavior identifies discrete scaling factors, at which integrate-and-fire behavior re-emerges in actual brains.** **(A)** The intuition for *Integration coefficient (IC)* - the degree, to which cross-correlogram of well-clustered brain's ensemble-spiking activity exceeds that of randomly clustered brain $P_{random}(\tau)$ is indicated *(shaded area)* and acts to quantify the neuron-like behavior of ensemble-nodes (neuronal clusters). See Methods for definition of *IC* and cross-correlograms. Shown for ensemble-spikes defined as **(top sub-panel)** two and **(bottom sub-panel)** four spikes. **(B)** *(Left)* Heatmap determines the ensemble-neuron size $N_N$ and ensemble-spike size $N_S$ used in re-scaling that best reproduces neuron-like behavior when clustered starting from a brain network. *(Right)* As a control, we show that the same signal integration behavior is lost when brain network is coarse-grained after a random shuffling (see Methods for procedure) that preserves strength of nodes (strength – sum of node's edge weights). This rules out the node strength sequence of brain network as the driver of its neuron-like behavior at the coarse scale. **(C)** The temporal re-scaling factor $N_t$ is obtained by calculating the auto-correlogram $P_{auto}(\tau)$ (see Methods for definition) and maximizing the ratio $P_{auto}(2)/P_{auto}(1)$ to reveal refractory period in coarse-grained network. **(D)** Local clustering coefficient among nodes within an ensemble-node corresponds with its integrate-and-fire behavior. Positive dependence indicates that highly interconnected set of nodes can join into an ensemble-node with more neuron-like integration of input.

Data Availability.

The data that support the findings of this study are available from the corresponding author through the Data Archive for the BRAIN Initiative (DABI) *Protecting the Aging Brain* directory.

Declaration of Interests

The authors declare no competing interests.

Acknowledgments

This research was funded by the National Science Foundation/BRAIN Initiative (ECCS1533257 HTS & LRMP), the Office of Naval Research (N00014-09-1-0069 HTS), the National Academies of Science & Engineering (NAKFICB8 LRMP), and the W.M. Keck Foundation (LRMP).

Author Contributions

Conceptualization, PT, HTS, AA, and LRMP; Investigation, AA; Writing—Original Draft, PT and AA; Writing—Critical Review & Editing, LRMP and HTS; Funding Acquisition, HTS and LRMP; Resources, HTS and LRMP; Supervision, HTS and LRMP.